\documentclass[pra,a4paper,twocolumn,english]{revtex4-1}
\usepackage{babel}
\usepackage{xcolor}
\usepackage{graphicx}
\usepackage{amsmath}
\usepackage{amssymb}
\usepackage{hhline}
\usepackage{bm}

\newcommand{\bc}{\begin{center}}
\newcommand{\ec}{\end{center}}
\newcommand{\be}{\begin{equation}}
\newcommand{\ee}{\end{equation}}
\newcommand{\bea}{\begin{eqnarray}}
\newcommand{\eea}{\end{eqnarray}}

\definecolor{darkblue}{rgb}{0.1,0.2,0.6}
\definecolor{darkred}{rgb}{0.8,0.1,0.2}
\usepackage[colorlinks,citecolor=darkblue,linkcolor=darkred,urlcolor=darkblue]
{hyperref} 
\usepackage[all]{hypcap} 

\newcommand{\ket}{\rangle}
\newcommand{\bra}{\langle}

\definecolor{commentcolor}{rgb}{0.1,0.2,0.6}
\definecolor{commentcolor2}{rgb}{1,0,0}
\definecolor{commentcolorF}{rgb}{0,0,1}
\definecolor{commentcolorD}{rgb}{1,0.1,.1}
\definecolor{todocolor}{rgb}{0.8,0.1,0.2}

\graphicspath {{Figures/}}

\begin{document}
\title{Restoring quasi-reversibility with a single topological charge}
\author{Juan Pablo \'Alvarez Z\'u\~niga}
\affiliation{Institut de Math\'ematiques de Toulouse, UPS, Toulouse, France}
\author{Romain Duboscq}
\email{romain.duboscq@math.univ-toulouse.fr}
\affiliation{Institut de Math\'ematiques de Toulouse, UPS, Toulouse, France}
\author{Juliette Billy}
\affiliation{Laboratoire Collisions, Agr\'egats et R\'eactivit\'e UPS, Toulouse, France}
\author{David Gu\'ery-Odelin}
\email{dgo@irsamc.ups-tlse.fr}
\affiliation{Laboratoire Collisions, Agr\'egats et R\'eactivit\'e UPS, Toulouse, France}
\author{Christophe Besse}
\email{christophe.besse@math.univ-toulouse.fr}
\affiliation{Institut de Math\'ematiques de Toulouse, UPS, Toulouse, France}

\date{\today}
\begin{abstract}
We numerically study a rotating Bose-Einstein condensate placed transiently over the critical rotation frequency i.e. in a regime where the rotation frequency is larger than the radial frequency of the confinement. 
We study the reversibility of this process depending on the strength of the interactions and the presence of vortices. We find that the reversibility is broken by the interactions in the absence of vortices but systematically quasi-restored in the presence of a single vortex. 
\end{abstract}

\maketitle

\numberwithin{equation}{section}


\section{Introduction}
The behavior of a rotating quantum fluid is known to exhibit some counterintuitive phenomena. For instance, by contrast with the rigid body rotation of a classical fluid, a quantum fluid reacts to the rotation, if sufficiently large, by nucleating vortices. This superfluid behavior was first observed in liquid He II \cite{heliumII}, and more recently revisited in the cold atom community with studies dedicated to rotating dilute Bose-Einstein Condensates (BECs) \cite{CornellImprinting,Dalibard1,ScienceKetterle,Dalibard2,Cornell1,Foot,Dalibard3,KetterleMAgneticField,Spielman}. 

Experimentally, various methods have been investigated to generate vortices using either phase imprinting \cite{CornellImprinting,KetterleMAgneticField,Spielman} or a rotating anisotropy superimposed to the confining potential. In this article, we concentrate on this latter technique for its versatility. This problem involves two frequencies: the rotation frequency, $\Omega$, and the trapping frequency, $\omega_0$, associated to the harmonic confinement. Vortices are here nucleated when the rotation resonantly excites quadrupole modes in the frequency domain $0 \le \Omega \le \omega_0/\sqrt{2}$ \cite{Dalibard1,ScienceKetterle,Cornell1,Foot} through a dynamical instability \cite{castin,lobo}. The critical rotation regime $\Omega\sim \omega_0$ has attracted a lot of attention \cite{a1,a2,a3,a4,a5,a6,a7,a8,a9,b1,b2,b3,b4,b5,b6,b7,b8,b9,b10}. In this regime and from a one-body point of view, the harmonic trapping force is exactly compensated by the centrifugal force. Thus atoms only  experience the Coriolis force in the rotating frame. This force is formally equivalent to the Lorentz force. The physics of neutral atoms in this regime is thus analogous to that of an electron gas in a uniform magnetic field. The ground energy level becomes macroscopically degenerated and phenomena related to the Quantum Hall effect with many vortices involved are expected \cite{b6}. 

In this article, we propose to explore the dynamics of a BEC in the presence or not of a single vortex in the rotation frequency domain $\Omega > \omega_0$. This regime is particularly difficult to study from an analytical point of view since there is no well adapted hydrodynamic formalism, no possible coarse graining approaches and no ground state. We therefore propose a numerical study of the corresponding out-of-equilibrium dynamics. The BEC is initially prepared at equilibrium with $\Omega < \omega_0$, and placed afterwards at a larger rotation frequency $\Omega > \omega_0$ either abruptly or adiabatically. As we shall discuss in the following, the change in $\Omega$ prohibits a perturbative treatment of the problem. Our main result is the observation of the restoring of a quasi-reversibility in such transformations, resulting from the presence of a single vortex. 

The paper is organized as follows. In section \ref{sec:Reminder}, we summarize the different regimes depending on the relative value of $\Omega$ and $\omega_0$. In Sec.~\ref{sec:num}, we detail our numerical procedures to prepare and place the BEC in the desired window of parameters. In Sec.~\ref{exact}, we provide a few exact analytical results in some limiting cases. Our numerical results are discussed in Sec.~\ref{sec:results}.


\section{Reminder on rotating particles}
\label{sec:Reminder}
We restrict our analysis to two dimensions i.e. in the plane perpendicular to the rotation axis. The potential experienced by the atom in the rotating frame $\mathcal{R}'$ is
\be
V_{e}(t,x',y')=\frac{1}{2}m\omega_0^2\left((1+\varepsilon(t))x^{\prime 2} +(1-\varepsilon(t))y^{\prime 2}  \right),
\label{potential}
\ee 
where $m$ is the particle mass and $\varepsilon$ accounts for the small rotating anisotropy. The corresponding classical equation of motions are time-independent~\cite{guery-odelin_spinning_2000}:
\be 
\left\{ \begin{aligned}
\ddot{x}^{\prime}(t)&= \left(-\omega_0^2(1+\varepsilon)+\Omega^2\right)x+2\Omega\dot{y}\\
\ddot{y}^{\prime}(t)&=  \left(-\omega_0^2(1-\varepsilon)+\Omega^2\right)y-2\Omega\dot{x}.
\end{aligned}
\right.
\label{eq:Single_Part_Eqs}
\ee
As expected, the confining potential frequency is reduced by the contribution of the centrifugal force: 
 $\omega_0^2(1\pm\varepsilon)\to\omega_0^2(1\pm\varepsilon-\Omega^2)$. 
The dispersion relation of these two linearly coupled equations yields stable solutions in two separated frequency domains: $\Omega<\Omega_-$ and
 $\Omega>\Omega_+$ with {$\Omega^2_\pm=\omega_0^2 (1\pm\varepsilon)$ and unstable solutions in the frequency range $\Omega_-<\Omega<\Omega_+$.  

For $\Omega_-<\Omega<\Omega_+$, the direction along the $y'$ axis is no more confining resulting into an instability of the particle. 
The observed stability for $\Omega>\Omega_+$ originates from the Coriolis force. This force favors a precession of the velocity which counteracts the 
repulsive force generated by the centrifugal force. We therefore obtain a dynamical stabilization of the atom in this regime. This is reminiscent of the magnetron stabilization in ion Penning traps~\cite{brown_geonium_1986}. 

The different regimes have been partially explored experimentally with an interacting Bose-Einstein condensate in the Thomas Fermi regime \cite{Dalibard2}. The routes to vortex nucleation through dynamical instabilities have been investigated in the range of parameter $0.5 < \Omega/\omega_0<1.1$ and $0 < \varepsilon < 0.03$ \cite{Dalibard2}. The instability window of rotation frequencies $\Omega_-<\Omega<\Omega_+$ has also been explored with a BEC from both a theoretical and an experimental point of view ($\Omega=\omega_0$ and $\varepsilon =0.09$) \cite{Dalibard3}. For this choice of parameters, the center of mass is unstable. The conclusions about the size of the cloud were somewhat counterintuitive: in the absence of interactions the cloud expands to infinity while it spirals out as a rigid body when repulsive interactions are sufficiently large \cite{Dalibard3}. The upper bound $\Omega_+(\varepsilon)$ is reduced in the presence of interactions while $\Omega_-(\varepsilon)$ is immune to the strength of interactions. In this article, we study the dynamics of a dilute interacting BEC in the regime $\Omega > \Omega_+(\varepsilon)$.


\section{Numerical procedure}
\label{sec:num}

Our results are based on the numerical resolution of the time-dependent 
2D Gross-Pitaevskii equation \cite{besse2004relaxation,GPELab} in the rotating frame associated to the potential (\ref{potential}): 
  \be
  \begin{aligned}
 i  \frac{\partial \Psi}{\partial t}=&-\frac{1}{2}\Delta\Psi+\frac{1}{2}\left((1+\varepsilon(t))x^{\prime 2} +(1-\varepsilon(t))y^{\prime 2}  \right)\Psi\\
 &+\beta|\Psi|^2\Psi-\Omega L'_z\Psi,
 \end{aligned}
  \label{eq:GPE_dimless}
  \ee
where the lengths are normalized to the harmonic length $a_0=(\hbar/m\omega_0)^{1/2}$ and the time and rotation frequency normalized to the angular frequency $\omega_0$. The parameter $\beta$ accounts for the strength of interactions.

For a given value of $\beta$ and $\Omega=0.9$ ($\varepsilon=0$), we determine the ground state wave function using an imaginary time evolution technique. The ground-state contains a number $N_v$ of vortices that depends on the interaction strength $\beta$~\cite{bao_ground_2005}. For $\beta=2$, the ground-state has no topological defect (i.e. no phase jump) and $N_v=0$ (see Fig.~\ref{fig:Groundstates}). For $\beta=5$, the ground-state accommodates a single vortex $N_v=1$ as explicitly shown on the phase map that exhibits a single phase singularity (see Fig.~\ref{fig:Groundstates}).  

\begin{figure}[htb]
\begin{center}
\includegraphics[width=0.98\columnwidth]{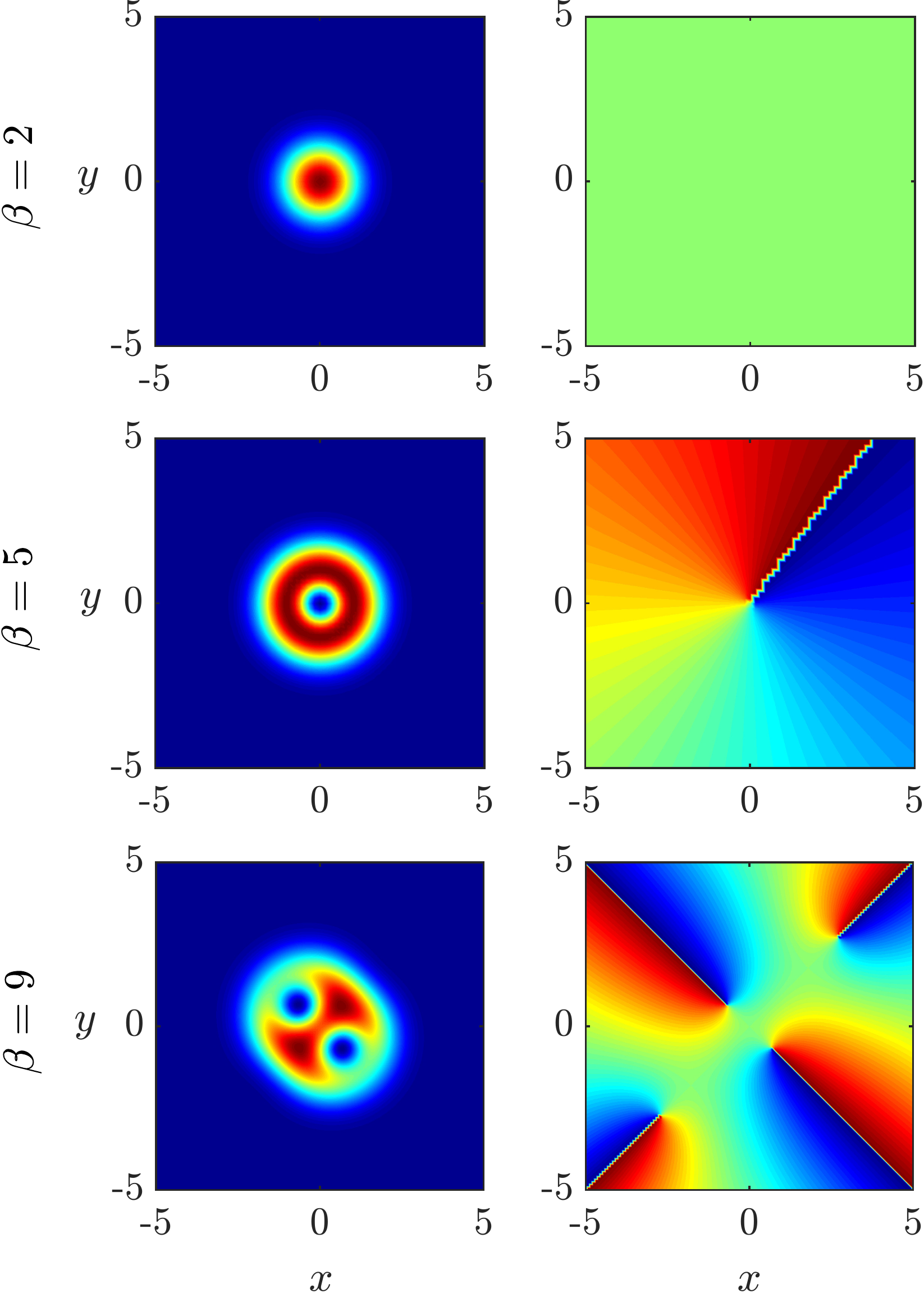}
\end{center}
\vspace{-6.5mm}
\caption{{\it Density profile (left) and phase map (right) of the ground-states for $\varepsilon=0$, $\Omega=0.9$ and $\beta=2$, $5$ and $9$.}}
\label{fig:Groundstates}
\end{figure}

Figure~\ref{fig:Procs} graphically summarizes the different procedures that we have investigated for the variations in time of $\varepsilon(t)$ and $\Omega(t)$.
The initial and final steps of the all procedures are always the same. We first ramp up the anisotropy $\varepsilon$ from 0 to a small 
value $\varepsilon_{max}$ of a few percent keeping the other parameters constant, and finally ramp it down symmetrically 
from  $\varepsilon_{max}$ to 0: 
\be
\varepsilon(t)= \left\{ \begin{aligned}
&\varepsilon_{\rm{max}}\times P(t),\, \text{(initial step)}\\
&\varepsilon_{\rm{max}}(1-P(t-2t_2-t_1)), \, \text{(final step)},\\
\end{aligned}
\right.
\label{eq:Epsilon_t}
\ee
where $2t_2$ is the amount of time over which the rotation frequency $\Omega$ is changed and $P(t)=6(t/t_1)^5-15(t/t_1)^4+10(t/t_1)^3$ is a smooth polynomial function increasing from 0 to 1 when $t$ spans the interval $[0;t_1]$. We fix $t_1=10\omega_0^{-1}$ to ensure a quasi-adiabatic ramping up of the anisotropy. The variation of the angular rotation frequency obeys 
\be
\Omega(t)= \left\{ \begin{aligned}
&0.9, t\in [0 ; t_1]\\ 
&0.9+(\Omega_{\rm{fin}}-0.9)f(t-t_1), t\in [t_1 ; t_1+t_2]\\ 
&\Omega_{\rm{fin}}-(\Omega_{\rm{fin}}-0.9)f(t-(t_1+t_2)) ,\\
&\qquad\qquad t\in [t_1+t_2 ; t_1+2t_2]\\
&0.9, t> t_1+2t_2. 
\end{aligned}
\right.
\label{eq:Omega_t}
\ee
with $f(t)=3(t/t_2)^2-2(t/t_2)^3$. 

We consider two different scenarii (see Fig.~\ref{fig:Procs}): (a) $\Omega$ is smoothly ramped up over $\Omega_+(\varepsilon)$ during the time interval $2t_2$ crossing the instability zone (interval $[\Omega_-(\varepsilon),\Omega_+(\varepsilon)]$) twice and (b) $\Omega$ is abruptly increased to a constant value
$\Omega_{\rm{fin}}>\Omega_+(\varepsilon)$, this plateau value being maintained during the time interval $2t_2$. The response of the system is analyzed as a function of $t_2$.

In practice, we keep track of the total energy defined by
\be
\begin{aligned}
E_{\beta,\,\Omega}(\Psi)=&\int\left[\frac{1}{2}|\nabla\Psi|^2 + V_{ext}|\Psi|^2 +\frac{\beta}{2}|\Psi|^4\right.\\
&\left.\qquad\quad-\Omega\Psi^{\ast}L_z\Psi \right]\,d^2r,
\end{aligned}
\label{eq:Energy}
\ee
the fidelity with respect to the initial ground-state $\Psi(t=0)=\Psi_0$, $\mathcal{F}(t)=|\langle \Psi_0|\Psi(t) \rangle|^2$ and the mean quadratic size of the cloud. 

\begin{figure}[!h]
\begin{center}
\includegraphics[width=0.98\columnwidth,clip]{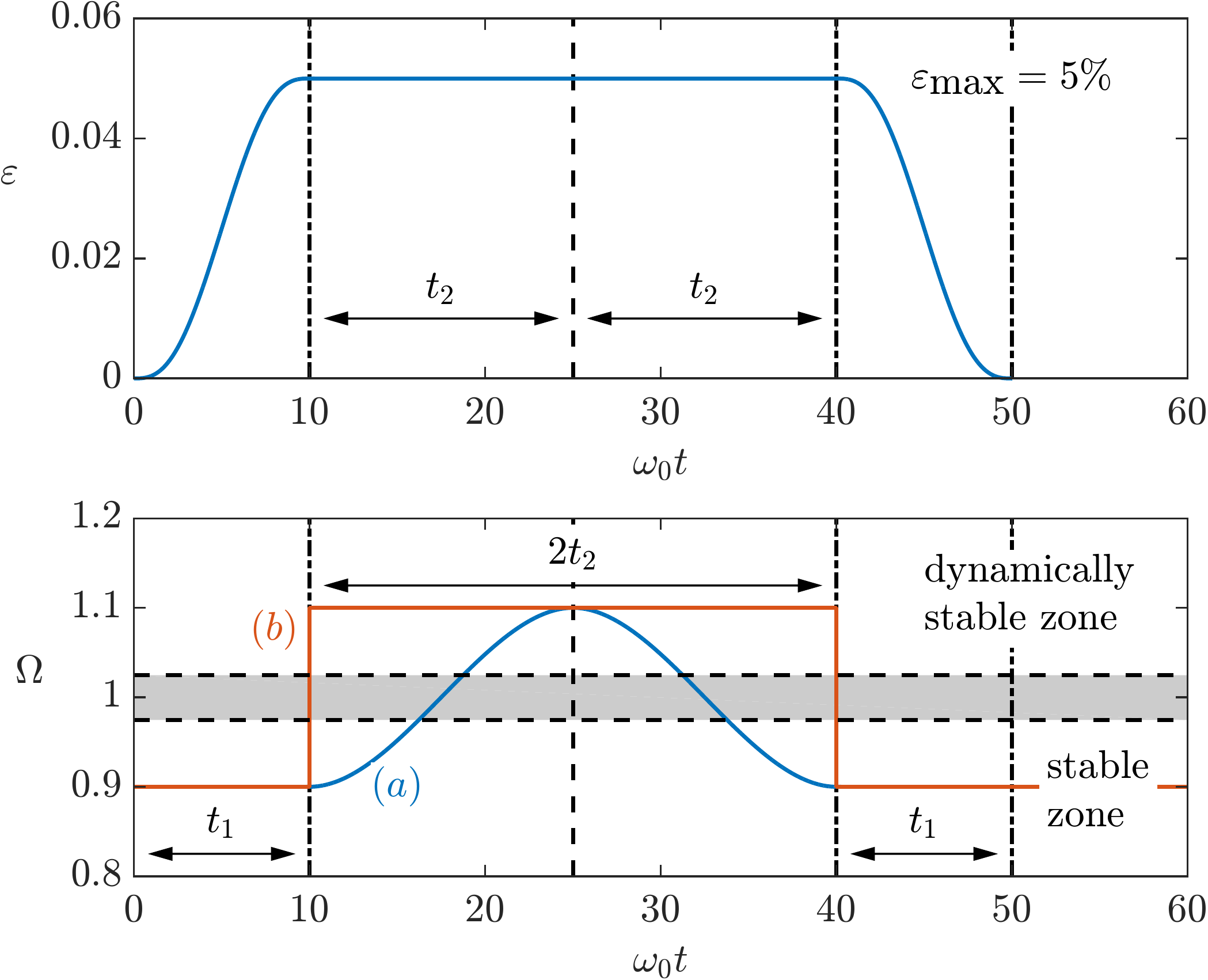}
\end{center}
\vspace{-6.5mm}
\caption{{\it Time evolution of the anisotropy $\varepsilon$ (top) and the rotation frequency $\Omega$ (bottom). The initial state is in the stable zone ($\Omega_i=0.9$). The shaded area depicts the instability zone in the absence of interactions $\Omega(t)\in[\sqrt{1-\varepsilon_{\rm{max}}};\sqrt{1+\varepsilon_{\rm{max}}}]$. To cross the instability zone, $\Omega$ is either abruptly changed (scenario (b)) or smoothly ramped (scenario (a)) to $\Omega_{\rm{fin}}$.}}
\label{fig:Procs}
\end{figure}
\vspace{-4.5mm}


\section{Exact results in limiting cases}
\label{exact}

To position the results obtained by our numerical simulations, it is instructive to work out analytically two limiting cases in the absence of vortices. 
We describe hereafter the evolution of the wave function through the exact determination of the time evolution of its mean quadratic size:
first in the absence of interactions and then in the opposite limit, the Thomas Fermi regime \cite{dgocct}.

\subsection{Non-interacting bosons}
\label{sec:Nv0_Class}

In the absence of interactions ($\beta=0$), we can infer the evolution of the size of the cloud using the Ehrenfest theorem with 
the time-dependent Hamiltonian
\be
H(t)=\frac{p_x^{'2}+p_y^{'2}}{2} +\frac{1}{2}\left(\left(1+\varepsilon\right)x^{\prime 2} +\left(1-\varepsilon\right)y^{\prime 2}  \right)-\Omega L'_z,
\label{eq:Quant_Ham_adim}
\ee
where $L'_z=x'\cdot p_y'-y'\cdot p_x'$. Any dynamical quantity $\chi_q$ which depends on the variables $x^{\prime}$, $y^{\prime}$, $p_x^{'}$ and $p_y^{'}$ has an average that evolves according to
\be
\frac{d\bra\chi_q\ket(t)}{dt}=i \bra\Psi(t)|\left[H(t),\chi_q \right]|\Psi(t)\ket.
\label{eq:Quant_Avg_eq}
\ee
This equation is exactly analogous to its classical counterpart based on the Boltzmann equation \cite{guery-odelin_collective_1999,guery-odelin_spinning_2000}.
The mean quadratic size involves the average quantity $\langle x^{\prime 2} +y^{\prime 2} \rangle$, and its time evolution involves other averages of quadratic  operators in $x^{\prime}$, $y^{\prime}$, $p_x^{'}$ and $p_y^{'}$. 
We eventually find that its evolution is given by a set of 10 linear equations (see appendix~\ref{sec:Apdx_Avg_Mom_Quant}) coupling the averages of the following quadratic operators:
$\chi_1=x^{\prime 2} +y^{\prime 2}$, $\chi_2=x^{\prime 2} -y^{\prime 2}$, $\chi_3=x'y'$, $\chi_4=x'p_x'+p_x'x'+y'p_y'+p_y'y'$, $\chi_5=x'p_x'+p_x'x'-y'p_y'-p_y'y'$, $\chi_6=x'p_y'+y'p_x'$, $\chi_7=x'p_y'-y'p_x'=L'_z$, $\chi_8=p_x^{\prime 2} +p_y^{\prime 2}$ , $\chi_9=p_x^{\prime 2} -p_y^{\prime 2}$ and $\chi_{10}=p_x'p_y'$. Interestingly, the total energy can also be expressed in terms of those averages  $\bra H(t)\ket  =\left(\bra\chi_8\ket+\bra\chi_1\ket+\varepsilon\bra\chi_2\ket \right)/2-\Omega\bra\chi_7\ket$.

For scenario (b), corresponding to a sudden change of $\Omega$, we solve numerically this set of equations with the initial conditions $\bra\chi_1\ket_0=\bra\chi_8\ket_0=1$ and $\bra\chi_i\ket_0= 0$ for $i\neq 1$ or $8$. In figure~\ref{fig:DeltaE_Class_vs_tpl_Jump}, we plot the relative energy difference between the initial and final states of the system $\Delta E/E_0=(E_f-E_0)/E_0$ as a function of $2t_2$, the duration over which $\Omega$ is changed. We observe an oscillatory behavior which can be readily explained as the selective excitation of a single eigenvalue of the $10\times 10$ matrix associated with the equations of motion. The corresponding frequency can be worked out 
\be
\omega_1=2 [1+\Omega^2_{\rm{fin}}-(\varepsilon^2_{\rm{max}} + 4\Omega^2_{\rm{fin}})^{1/2}]^{1/2},
\label{eq:Eigen_Freqs}
\ee
which, for $\varepsilon_{\rm{max}}=5\%,$ yields the oscillation periods $T_1\approx 32.3482$ for $\Omega_{\rm{fin}}=1.1$ and $T_1\approx 15.8112$ for $\Omega_{\rm{fin}}=1.2$ in perfect agreement with the observed periods. As a matter of fact, the amplitude of oscillations of the relative energy difference decreases when $\Omega_{\rm fin}$ increases ($7.5\%$ for $\Omega_{\rm{fin}}=1.1$ and $4\%$ for $\Omega_{\rm{fin}}=1.2$). The jump in $\Omega$ therefore selects a single eigenvalue which garanties the reversibility of the process i.e. the periodic cancelation of the relative energy difference, $\Delta E/E_0$, as a function of time. We have checked that applying the same procedure with a different initial rotation frequency ($\Omega_i=0.5$), we  recover the same behavior (see figure ~\ref{fig:DeltaE_Class_vs_tpl_Jump}). In the following we choose $\Omega_i=0.9$.

\begin{figure}[!h]
\begin{center}
\includegraphics[width=0.98\columnwidth,clip]{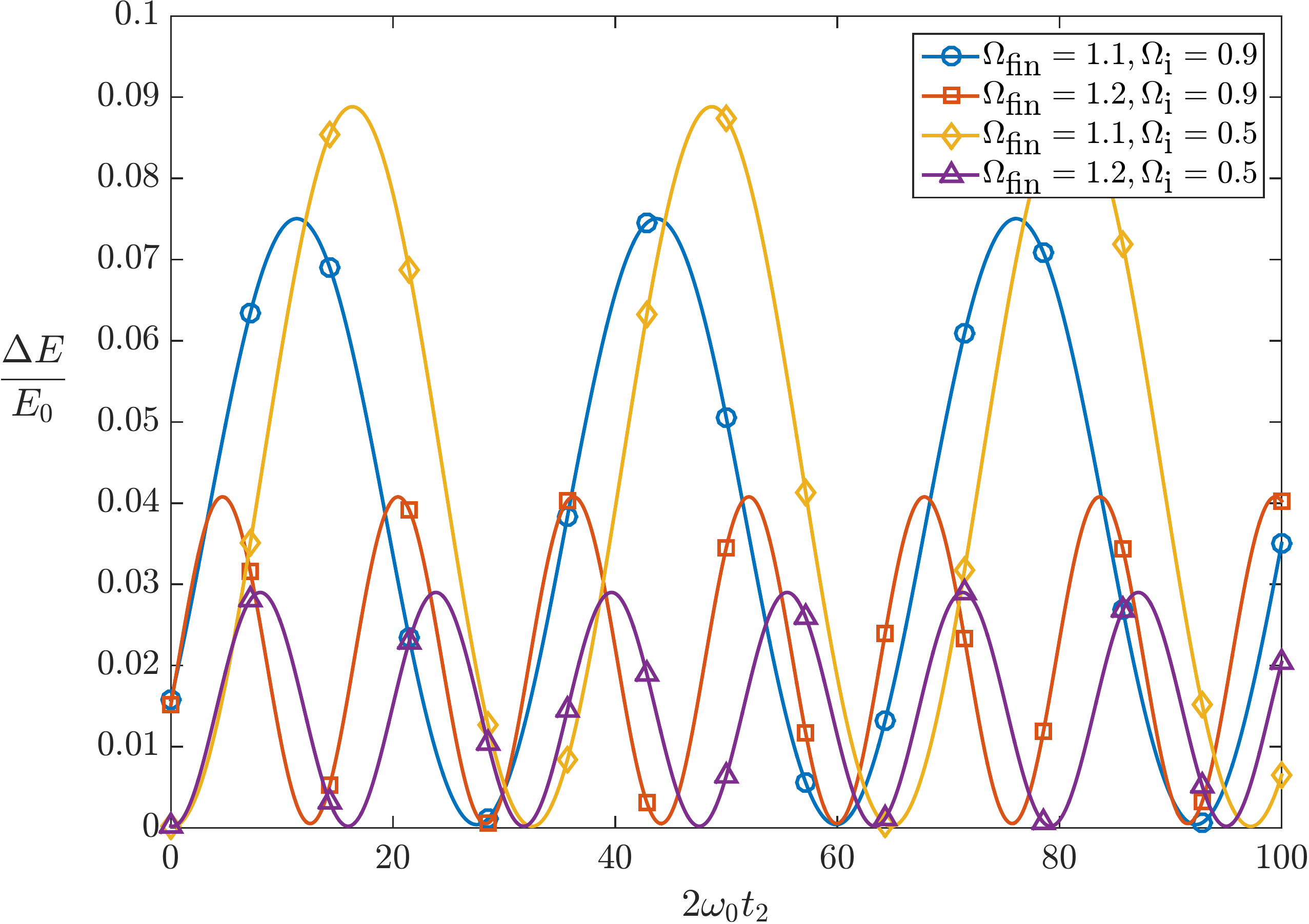}
\end{center}
\vspace{-7mm}
\caption{{\it Relative energy difference between the initial and final states of the system $\Delta E/E_0=(E_f-E_0)/E_0$ as a function of $2\omega_0 t_{2}$, applying scenario (b) to a non-interacting BEC ($\beta=0$) with $\varepsilon_{\rm{max}}=5\%$, and for various values of $\Omega_i$ and $\Omega_{\rm{fin}}$.}}
\label{fig:DeltaE_Class_vs_tpl_Jump}
\vspace{-2mm}
\end{figure}


\subsection{The Thomas Fermi limit}

For sufficiently large interaction strength $\beta$ and in the absence of vortices, the 2D GPE equation is equivalent in the corotating frame to a set of two hydrodynamic-like equations associated to the phase and modulus of the wave function $\Psi(x,y;t)=\rho^{1/2}(x,y;t)e^{i\theta(x,y;t)}$. The continuity equation reads
\be
\frac{\partial \rho}{\partial t}+\boldsymbol{\nabla}\left[\rho\left(\boldsymbol{v}-\boldsymbol{\Omega}\times\boldsymbol{r} \right) \right]=0,
\label{continuity}
\ee
with $\boldsymbol{v}=\boldsymbol{\nabla}\theta$, and the Euler-like equation is given by
\be
\begin{aligned}
\frac{\partial \boldsymbol{v}}{\partial t}+\boldsymbol{\nabla}&\left[\frac{\boldsymbol{v}^2}{2}+\frac{1}{2}\left((1+\varepsilon)x^2+(1-\varepsilon)y^2 \right)\right.\\
&\quad\left.+\beta\rho-\boldsymbol{v}\cdot(\boldsymbol{\Omega}\times\boldsymbol{r}) \right]=0.
\end{aligned}
\label{Euler}
\ee
Equations (\ref{continuity}) and (\ref{Euler}) are easily solved using the ansatz 
\begin{eqnarray}
\rho(x,y;t)  & = & a_0 + a_x x^2 + a_y y^2+a_{xy}xy \nonumber \\
\theta(x,y;t) & = &  \alpha_x x^2/2 + \alpha_y y^2/2+\eta xy
\label{ansatz}
\end{eqnarray}
where $a_0$, $a_x$, $a_y$, $a_{xy}$, $\alpha_x$, $\alpha_y$ and $\eta$ are time-dependent variables. We find a closed set of non-linear coupled equations for these variables (see Appendix \ref{sec:Apdx_nlsolhydro}) which provides a non-linear oscillation of period $T_e\simeq 13.38$ for $\varepsilon_{\rm{max}} = 0.05$, $\Omega_i=0.9$, $\Omega_{\rm fin}=1.1$ and whose value does not depend on $\beta$.


\section{Numerical results outside the limiting cases}
\label{sec:results}

This section first summarizes the results we have obtained for a jump of the rotation frequency $\Omega$ (scenario (b)) at a finite value of the interaction strength parameter $\beta$ in the absence and in the presence of vortices. We also report on the results obtained with a smooth variation of the rotation frequency (scenario (a)).

\subsection{Results for a sudden variation of $\Omega$}

We consider a ground-state with no topological charge ($N_v=0$, $0<\beta<4$) and change suddenly the rotation frequency from $\Omega_i=0.9$ to $\Omega_{\rm{fin}}=1.1$, for an anisotropic parameter $\varepsilon_{\rm{max}}=5\%$. As previously, we plot the relative energy difference between the initial and final states as a function of $2\omega_0 t_2$. We find three main differences compared to the case without interactions (see Fig.~\ref{fig:DeltaE_Jump_No_vort_Betas}): oscillations have a lower frequency; they are slightly damped and their relative amplitude is dramatically reduced  (amplitude of $\sim 0.5\%$ (for $\beta=2$) to be compared to $7.5\%$ for $\beta=0$). This latter feature results from the large contribution of the interaction energy to the total energy. The slight damping suggests that, in the presence of interactions, many modes are contaminated by the excitation process. In the absence of interactions, we have seen that the breathing mode (i.e. $\chi_1$) is coupled to the quadrupole mode (i.e. $\chi_2$). It is known that the frequency of both modes decreases as the interaction strength $\beta$ increases \cite{dgocct}. We recover here the same tendency in the corotating frame.

\begin{figure}[!h]
\begin{center}
\includegraphics[width=0.98\columnwidth,clip]{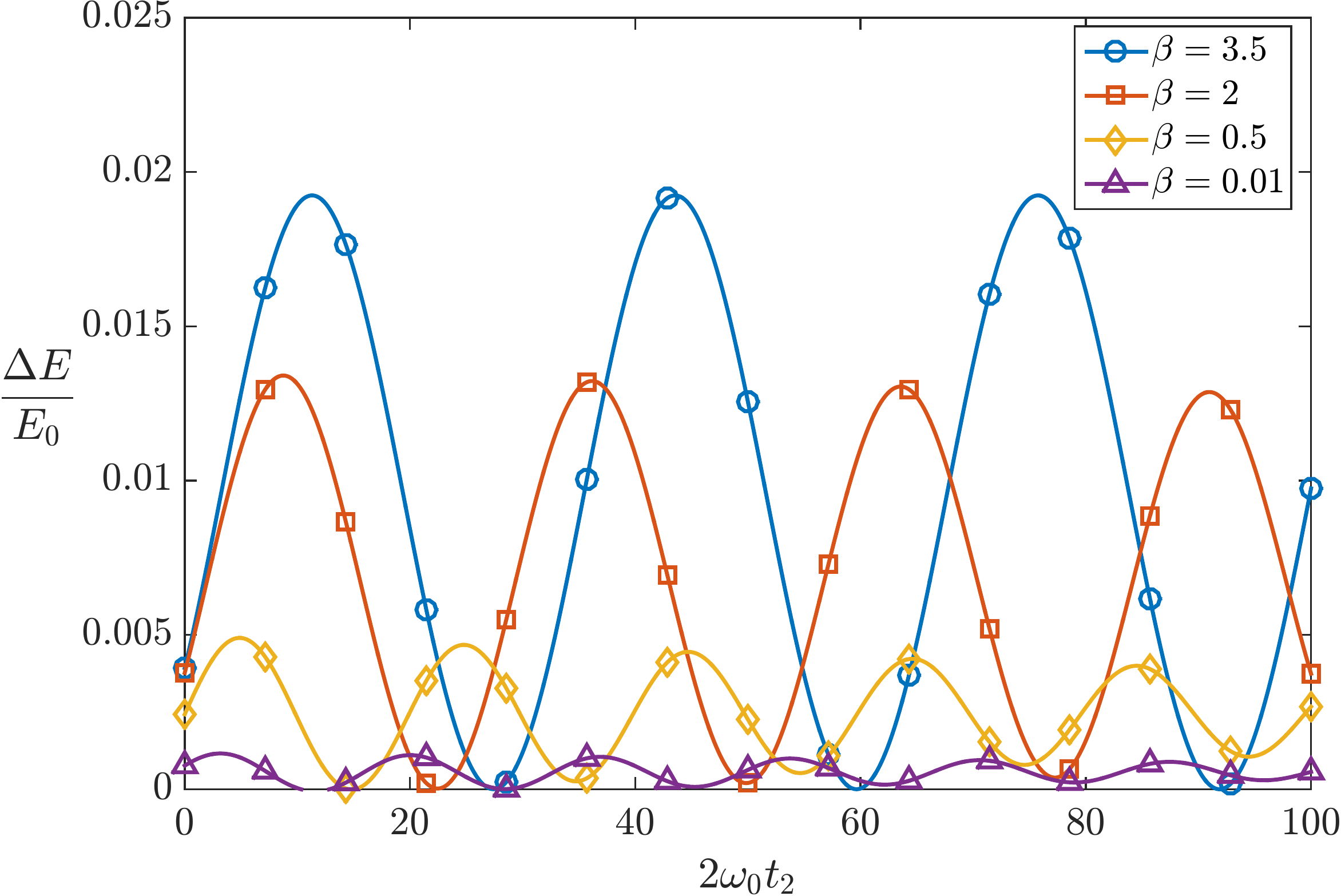}
\end{center}
\vspace{-7mm}
\caption{{\it Relative energy difference between the initial and final states $\Delta E/E_0$ as a function of $2\omega_0 t_2$ applying the scenario $(b)$ (with $\varepsilon_{\rm{max}}=5\%$, $\Omega_i=0.9$ and $\Omega_{\rm{fin}}=1.1$) to ground-states without vortices for different interaction strength $\beta$.}}
\label{fig:DeltaE_Jump_No_vort_Betas}
\end{figure}

Applying the very same procedure for different interaction strength $\beta$ in the interval $4<\beta<8.5$ i.e.~in the presence of a single vortex, we observe an oscillation that is not damped and whose period is close to that for $\beta=0$ at the lowest value of $\beta$ for which a single vortex appears, and that decreases with $\beta$.
Figure \ref{fig:Jump_finie} provides for $\beta=5$ the evolution of the relative energy difference between the initial and final state along with the fidelity of the final state with respect to the initial one for various values of the anisotropy parameter $\varepsilon_{\rm{max}}$.

\begin{figure}[!t]
\begin{center}
\includegraphics[width=0.98\columnwidth,clip]{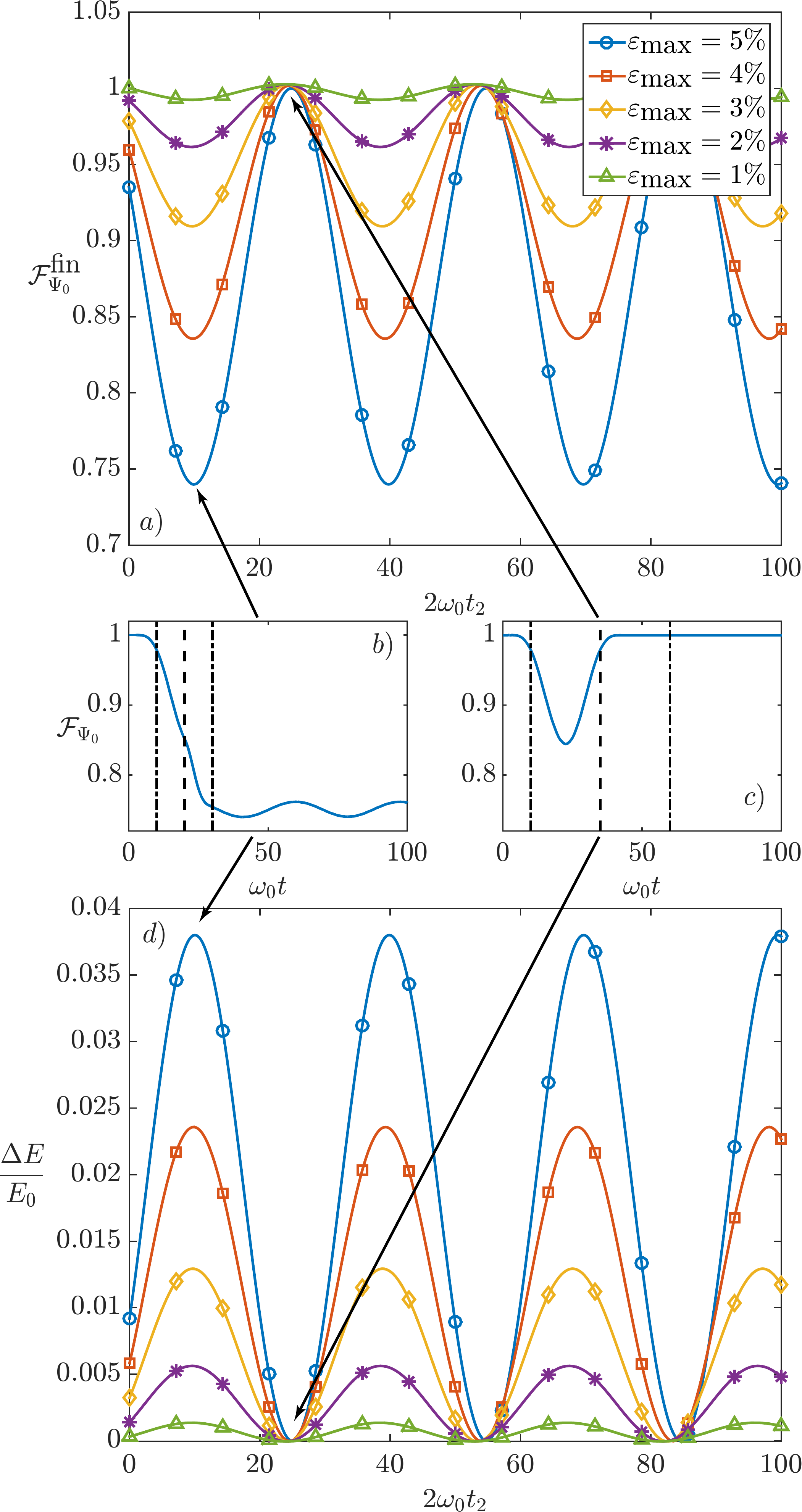}
\end{center}
\vspace{-7mm}
\caption{{\it a) Fidelity of the final state with respect to the initial one as a fonction of $2\omega_0 t_2$ applying scenario (b) (with $\Omega_i=0.9$ and $\Omega_{\rm{fin}}=1.1$) to a BEC with a single vortex ($\beta=5$) for various values of the anisotropic parameter $\varepsilon_{\rm{max}}$. b) and c) Evolution of the fidelity over time respectively at the first minimum and the first maximum of the final fidelity. d) Relative energy difference between the initial and final states as a function of $2 \omega_0  t_2$.}}
\label{fig:Jump_finie}
\end{figure}

Our numerical results for a sudden change in the rotation frequency and for $0<\beta<10$ are summarized in Fig.~\ref{fig:Period_vs_Betas}. We plot the period of oscillation of the relative energy difference $\Delta E/E_0$ as a function of $\beta$ and compare it to the predictions of Sec.~\ref{exact} in the absence of interactions and in the Thomas-Fermi regime respectively. The data obtained in the absence of vortices are in between those two limiting cases. We also plot on the same figure the contrast $C_{st}$ of the excess of energy $\Delta E/E_0$ as a function of $\beta$ \cite{contrast}. Remarkably, the contrast is restored to unity \cite{footnote2} only in the window of interaction strength that corresponds to the presence of a single vortex. In the presence of two vortices, the contrast drops again drastically, which is probably related to the rotation symmetry break.

\begin{figure}[!h]
\begin{center}
\includegraphics[width=0.98\columnwidth,clip]{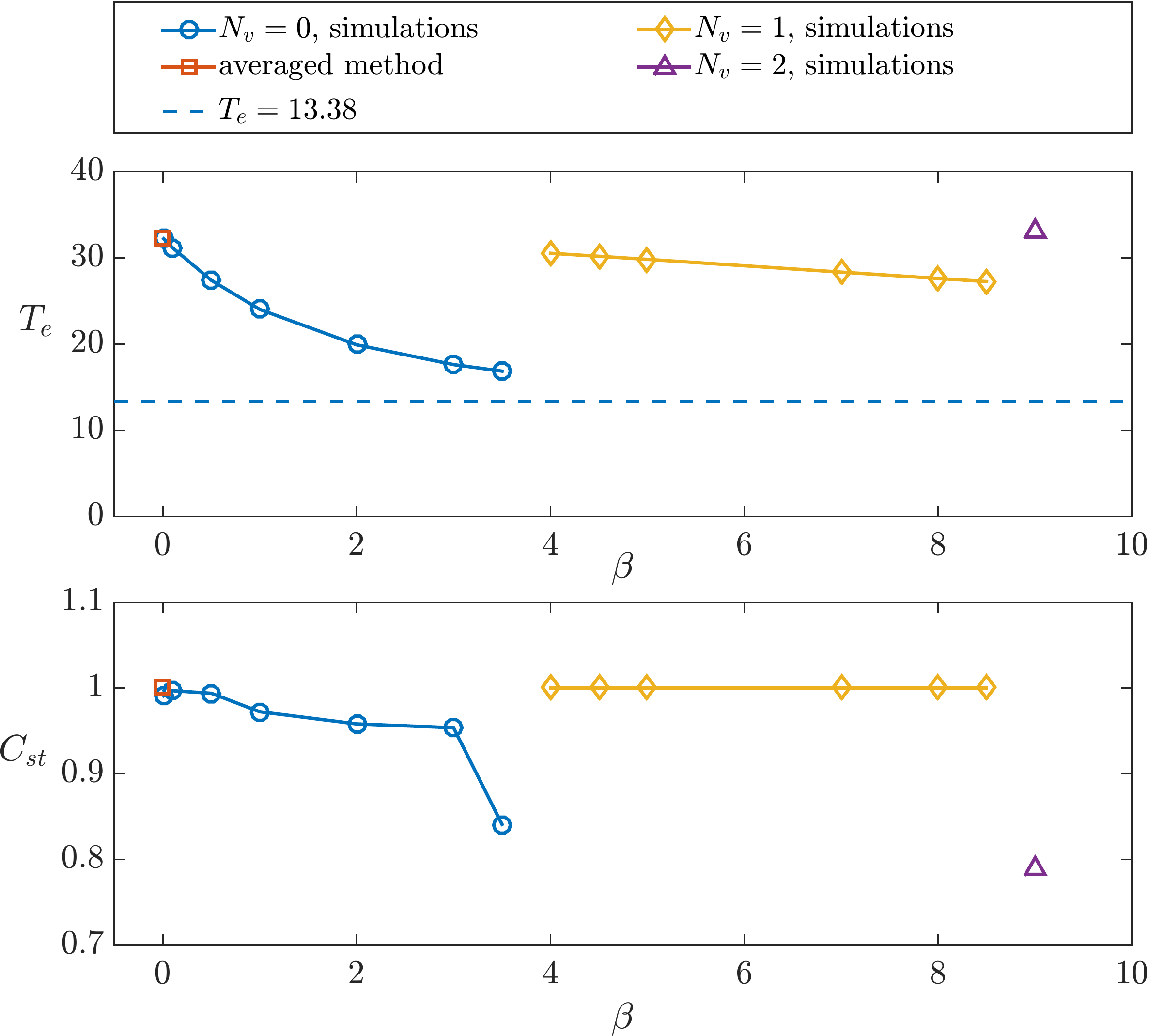}
\end{center}
\vspace{-7mm}
\caption{{\it Period (top) and contrast $C_{st}$ (bottom) of the excess of energy $\Delta E/E_0$ as a function of the interaction strength $\beta$ when applying scenario (b) (with $\varepsilon_{\rm{max}}=5\%$, $\Omega_i=0.9$ and $\Omega_{\rm{fin}}=1.1$) to different ground-states with $N_v=0$ (circles), $N_v=1$ (diamonds), and $N_v=2$ (triangle) vortices. The square corresponds to the prediction of Sec. IV in the non-interacting case and the lower dashed line to the one in the Thomas-Fermi regime.}}
\label{fig:Period_vs_Betas}
\end{figure}

We have also study the dependence of the amplitude, $b_e$, and the period, $T_e$, of the oscillations of $\Delta E/E_0$ with the final rotation frequency, $\Omega_{\rm fin}$ (see Fig.~\ref{fig:Jump_Coeffs_Om_fins}). For $\beta=0$, we observe a divergence of both the amplitude and the period as we approach the instability zone $\Omega \longrightarrow \Omega_+(\varepsilon)$. The same behavior is observed for $\beta=5$. However, the divergence in the amplitude is less pronounced. This is due to the fact that $\Omega_+(\varepsilon)$ is renormalized by the interactions as explained in \cite{Dalibard3}.

\subsection{Smooth variation of the rotation frequency}

In this section, we consider a smooth variation of the rotation frequency from its initial value $\Omega_i$ to its final value $\Omega_{\rm fin}$ (scenario (a)). 
As a direct consequence, the rotation frequency crosses the instability region $[\Omega_-,\Omega_+]$ twice. In the absence of interactions, the variations of the relative excess of energy, $\Delta E/E_0$, as a function of the time $t_2$ are given in  Fig.~\ref{fig:DeltaE_Class_vs_t2} for various values of the anisotropic parameter $\varepsilon_{\rm{max}}$. In the instability region, the cloud size explodes; this is the reason why $\Delta E/E_0$ increases with $t_2$ i.e. with the time spent in this instability window. Remarkably, this instability does not prohibit the quasi-reversibility of the process, and we find discrete values of time $t_2$ for which the excess of energy cancels out. These "magic" time durations are, for $\varepsilon_{\rm{max}}=5\%$, $2\omega_0t_2\approx 86$ 
and $2\omega_0t_2\approx 184$ with the respective relative energy differences $2\cdot 10^{-4}$ and $2\cdot 10^{-6}$.  

\begin{figure}[!h]
\begin{center}
\includegraphics[width=0.98\columnwidth,clip]{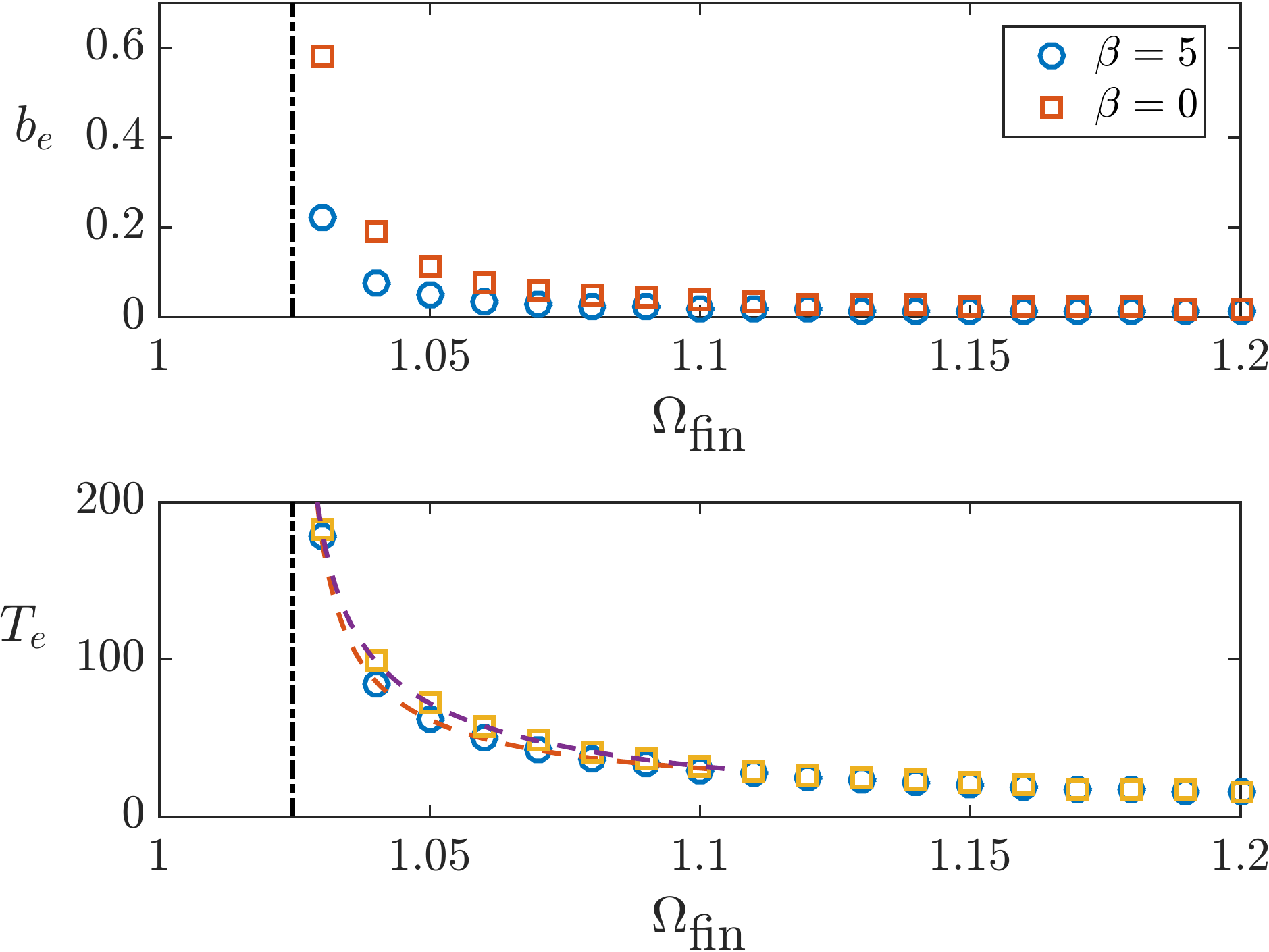}
\end{center}
\vspace{-7mm}
\caption{{\it Amplitude $b_e$ and period $T_e$ of the excess of energy $\Delta E/E_0$ with the maximum rotation frequency, $\Omega_{\rm fin}$ in the single vortex case ($\beta=5$) and in the non-interacting case ($\beta=0$), following the scenario (b) with $\Omega_i=0.9$ and $\varepsilon_{\rm{max}}=5\%$. The vertical dashed line shows the upper limit of the instability zone $\sqrt{1+\varepsilon_{\rm{max}}}$ in the absence of interactions.}}
\label{fig:Jump_Coeffs_Om_fins}
\end{figure}

\begin{figure}[!h]
\begin{center}
\includegraphics[width=0.98\columnwidth,clip]{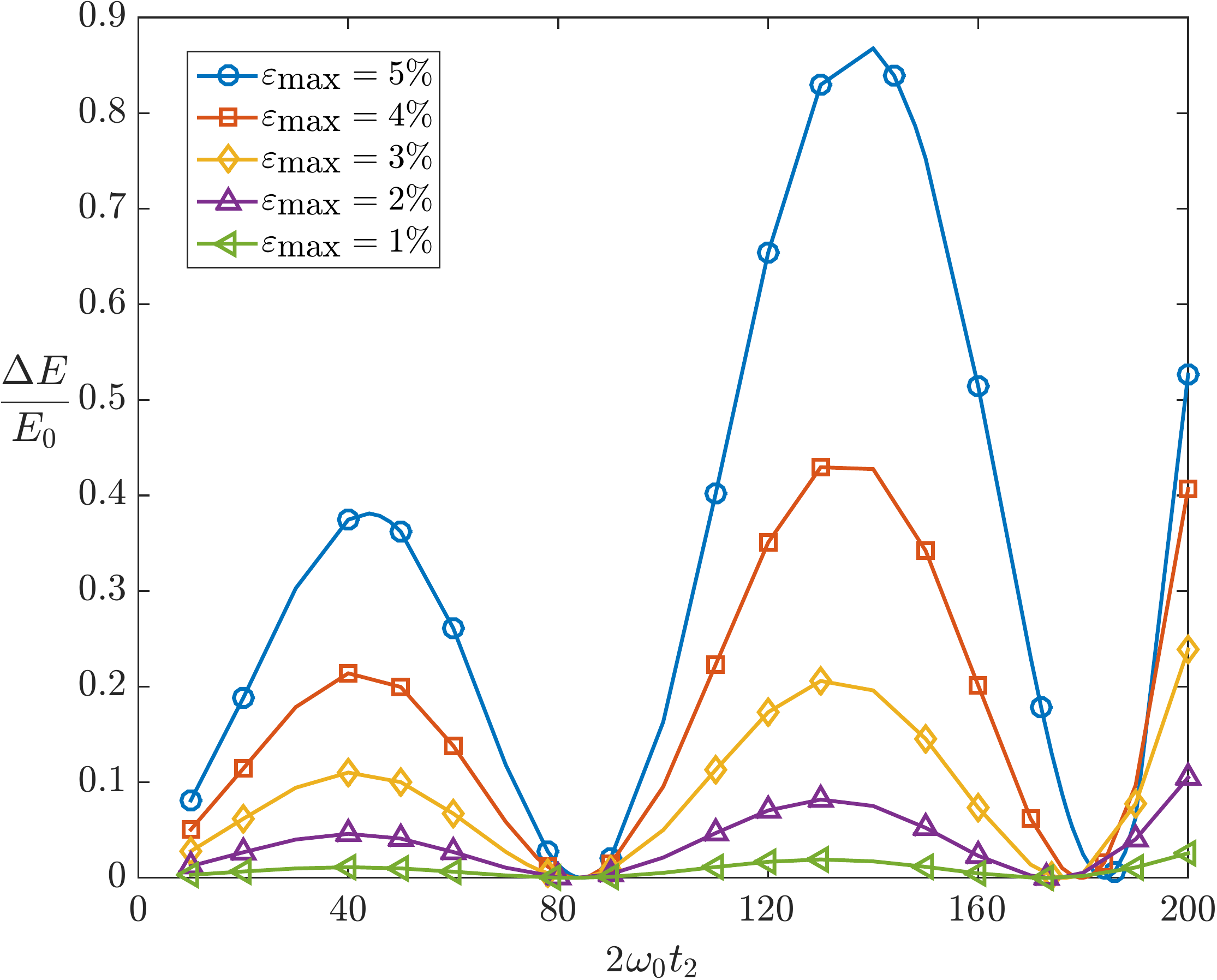}
\end{center}
\vspace{-7mm}
\caption{{\it Relative energy difference $\Delta E/E_0$ as a function of $2\omega_0 t_{2}$ in the non-interacting case ($\beta=0$), using scenario (a) (with $\Omega_i=0.9$ and $\Omega_{\rm{fin}}=1.1$) for various values of the anisotropic parameter $\varepsilon_{\rm{max}}$. }}
\label{fig:DeltaE_Class_vs_t2}
\end{figure}

This is to be contrasted with our observations at finite interaction strength $\beta$. As an example, we start from a ground-state without vortices ($\beta=2$). The relative energy difference is plotted in figure~\ref{fig:DeltaE_vs_t2_no_vortex} : there are two clear local minima for $2\omega_0t_2\approx 34$ ($\Delta E/E_0\approx0.49\%$) 
and $2\omega_0t_2\approx 174$ ($\Delta E/E_0\approx16\%$). In the absence of vortices initially ($\beta < 4$), the wave function has its phase strongly affected by crossing the instability region as it can be seen for instance in Fig.~\ref{fig:DeltaE_vs_t2_no_vortex}, where we provide the final density and phase profiles of the wave-function at the local minimum of the relative energy difference at $2\omega_0t_2=174$. 
The transient entrance of vortices in the course of the out of equilibrium dynamics breaks the quasi-reversibility. 
A fingerprint of the quasi-reversibility observed in the absence of interactions remains with the presence of a
local minimum at $2\omega_0t_2\approx 174$ but at a non zero value of the relative energy difference.

\begin{figure}[!h]
\begin{center}
\includegraphics[width=0.96\columnwidth,clip]{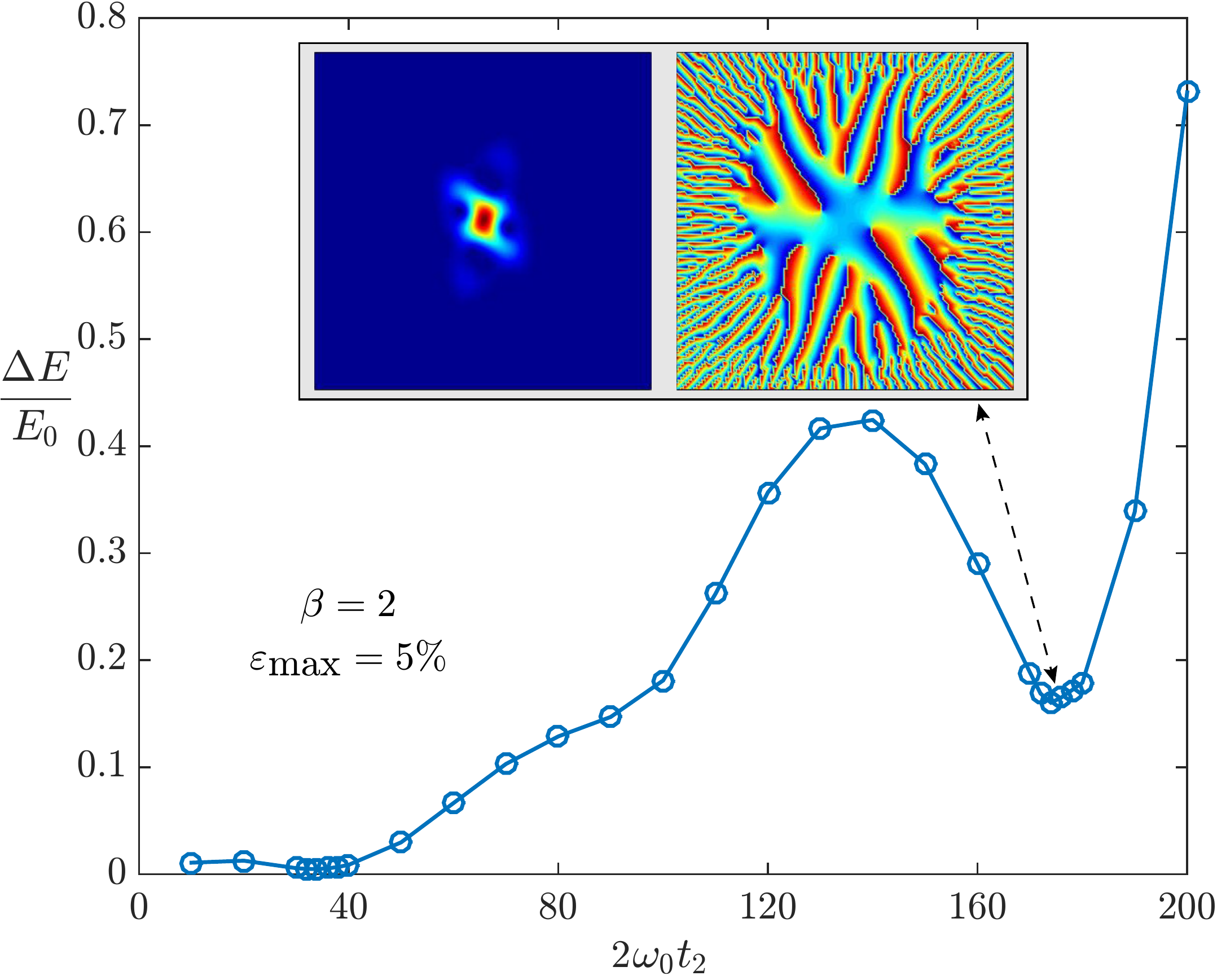}
\end{center}
\vspace{-7mm}
\caption{{\it Relative energy difference as a function of $2\omega_0 t_2$ when applying scenario (a) to an interacting BEC without vortices: $\beta=2$, $\varepsilon_{\rm{max}}=5\%$, $\Omega_i=0.9$ and $\Omega_{\rm{fin}}=1.1$. The insets show the final density and phase profiles after the instability sweep at the local minimum at $2\omega_0t_2=174$.}}
\label{fig:DeltaE_vs_t2_no_vortex}
\end{figure}

\begin{figure}[!h]
\begin{center}
\includegraphics[width=0.98\columnwidth]{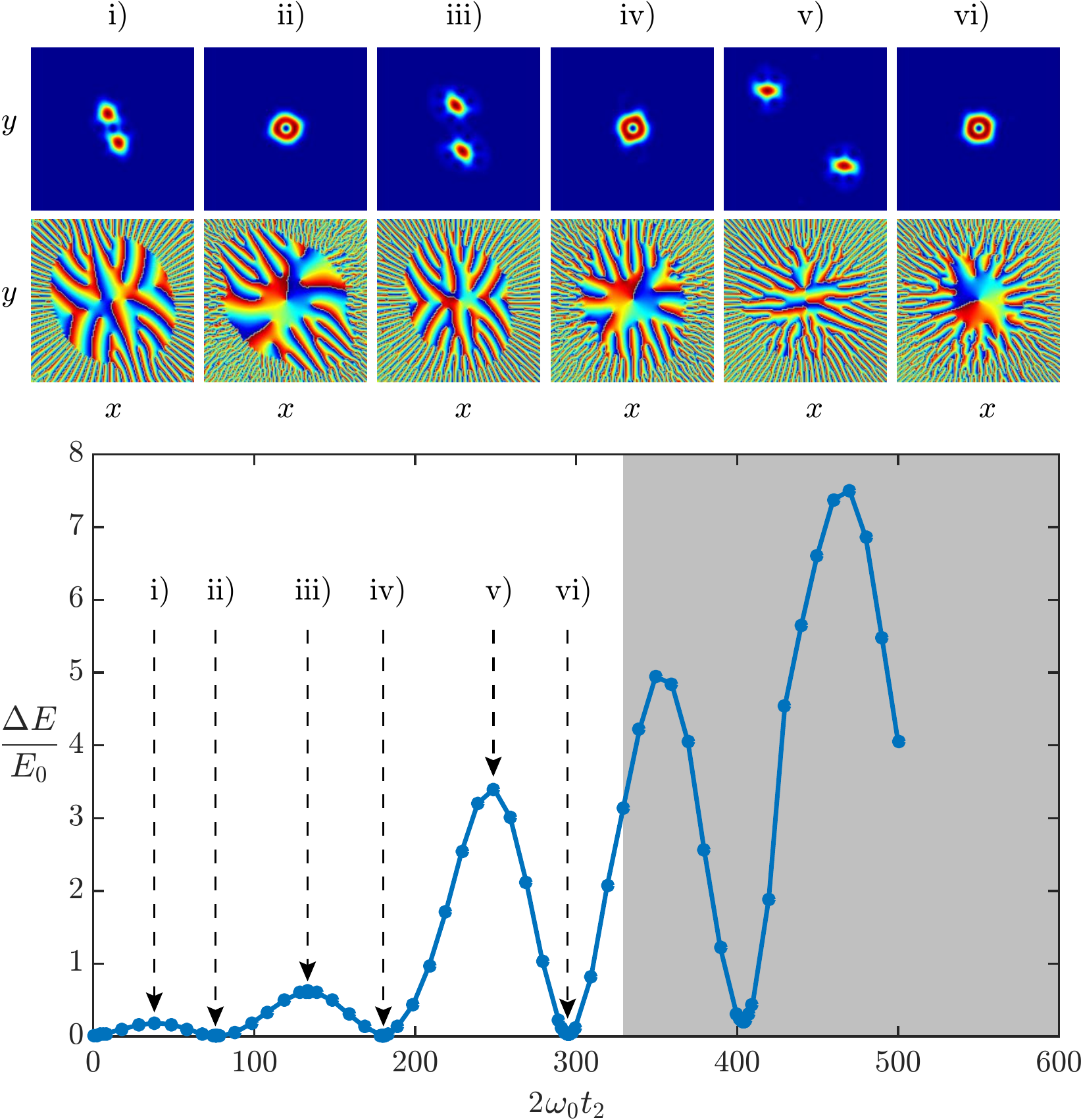}
\end{center}
\vspace{-7mm}
\caption{{\it Relative energy difference as a function of $2\omega_0 t_2$ when applying scenario (a) to a BEC with a single vortex ($\beta=5$) for $\Omega_i=0.9$, $\Omega_{\rm{fin}}=1.1$ and $\varepsilon_{\rm{max}}=5\%$. The shaded area depicts the zone where the accuracy of the results is slightly altered by the finite size of the computation grid. The density (phase) profiles at the extrema denoted by the roman numbers are shown in the top (bottom) strips.}}
\label{fig:DeltaE_vs_t2_inst_1_vortex_eps_5}
\end{figure}

In contrast, the same scenario applied to a BEC that contains a single vortex ($\beta=5$) for various values of $t_2$ at 
fixed $\Omega_{\rm{fin}}=1.1$ and $\varepsilon_{\rm{max}}=5\%$ restores a curve reminiscent of that without interactions with well pronounced minima (see Fig.~\ref{fig:DeltaE_vs_t2_inst_1_vortex_eps_5}). For such minima, the energy given to the system while
sweeping upwards the instability zone is almost exactly cancelled out during the downward sweep. 
We have checked that an adiabatic sweep towards lower rotation frequency values (with $\Omega_{\rm{fin}}=0.7$) has no impact on the relative energy.  

The roman numbers in fig.~\ref{fig:DeltaE_vs_t2_inst_1_vortex_eps_5} correspond to the extrema of the relative energy difference. We show the associated density and phase profiles. 
As expected, the points with almost vanishing energy difference have a 
final density profile very close to that of the initial single vortex ground-state, while the maxima present a final cloud almost separated 
in two sub-clouds, linked by a central elongated low density part. The more elongated the cloud, the more energy it has. However, the central 
vortex present in the initial ground-state remains present as confirmed by the computation of the circulation around the center. The robustness of the central topological defect 
is confirmed in this context since the explosion of the cloud resulting from the crossing of the instability region does not change the circulation. 

\section{Conclusion}

In this article, we have explored numerically a singular regime for rotating BEC that cannot be accessed in a simple manner by analytical means. We have observed the strong influence of the presence of a single vortex when the BEC is driven over the critical rotation frequency. The time reversibility observed in the absence of interactions can be explained by classical equations. Numerically we observed that this effect is destroyed when repulsive interactions are increased but restored in the regime for which interactions are sufficiently large so that the ground state accommodates a single vortex. Intensive numerical experimentations suggest that the quasi-reversibility that we have highlighted is not impacted by the numerical parameters (grid size and timestep). It is well known that the moment of inertia of a BEC with many vortices tends to the classical value. Here, we observe on the dynamics of the breathing mode that a single vortex restores a classical dynamics while more vortices would not. 

We have studied the quasi-reversibility in the presence of a harmonic trap. In \cite{Proukakis1,Proukakis2}, the authors emphasize a phenomenon which suggests that the choice of an anharmonic potential could lead to an instability of a topological defect through the emission of sound. We thus led additional numerical experiments in the presence of a quartic perturbation for our 2D potential revealing that the quasi-reversibility is still present. This suggests that this process is robust with respect to the choice of the potential, and therefore can be safely investigated experimentally.
\\
\noindent\textbf{Acknowledgements}

This work was partially supported by the French ANR grant ANR-12-MONU-0007-02 BECASIM ("Mod\`eles Num\'eriques" call).

\appendix 


%

\section{Average method for quadrupolar moments}
\label{sec:Apdx_Avg_Mom_Quant}

Using equations (\ref{eq:Quant_Avg_eq}) for the Hamiltonian (\ref{eq:Quant_Ham_adim}), we find the following set of coupled equations:
 \be
\begin{aligned}
\frac{d\bra \chi_1\ket}{dt}-\bra \chi_4\ket=&0,\\
\frac{d\bra \chi_2\ket}{dt}-\bra \chi_5\ket-4\Omega\bra \chi_3\ket=&0,\\
\frac{d\bra \chi_3\ket}{dt}-\bra \chi_6\ket+\Omega\bra \chi_2\ket=&0,\\
\frac{d\bra \chi_4\ket}{dt}-2\bra \chi_8\ket+2\bra \chi_1 \ket+2\varepsilon\bra \chi_2\ket=&0,\\
\frac{d\bra \chi_5\ket}{dt}-2\bra \chi_9\ket+2\bra \chi_2 \ket+2\varepsilon\bra \chi_1\ket -4\Omega(t)\bra \chi_6\ket=&0,\\
\frac{d\bra \chi_6\ket}{dt}-2\bra \chi_{10}\ket+2\bra \chi_3\ket+\Omega\bra \chi_5\ket=&0,\\
\frac{d\bra \chi_7\ket}{dt}-2\varepsilon\bra \chi_3\ket=&0,\\
\frac{d\bra \chi_8\ket}{dt}+\bra \chi_4\ket+\varepsilon\bra \chi_5\ket=&0,\\
\frac{d\bra \chi_9\ket}{dt}-4\Omega\bra \chi_{10}\ket+\bra \chi_5\ket+\varepsilon\bra \chi_4\ket=&0,\\
\frac{d\bra \chi_{10}\ket}{dt}+\Omega\bra \chi_9\ket+\bra \chi_6\ket+\varepsilon\bra \chi_7\ket=&0.\\
\end{aligned}
\label{Apdx_eq:Set_Eq_Quant}
\ee
We can readily check the conservation of the total energy in the case of time-independent rotation frequency $\Omega$ and anisotropy $\varepsilon$, $d\bra H(t)\ket/dt=0$.

\section{Nonlinear solution of the hydrodynamic equations}
\label{sec:Apdx_nlsolhydro}

The stationary solutions of Eqs.~(\ref{continuity}) and (\ref{Euler}) reads \cite{recati_overcritical_2001}
$\boldsymbol{v}_{\rm st}=\eta_0\boldsymbol{\nabla}(xy)$ and $\widetilde{\rho}_{\rm st}=(\widetilde{\mu}/\beta)\left(1-x^2/R_x^2-y^2/R_y^2\right)$
with $\widetilde{\omega}_x^2=(1+\varepsilon)+\eta_0^2-2\eta_0\Omega$, $\widetilde{\omega}_y^2=(1-\varepsilon)+\eta_0^2+2\eta_0\Omega$, 
$R_x^2=2\widetilde{\mu}/\widetilde{\omega}_x^2$, and $R_y^2=2\widetilde{\mu}/\widetilde{\omega}_y^2$.
The constant $\widetilde{\mu}$ is the chemical potential. Its value is determined through the normalization to unity of the density $\widetilde{\mu}^2=\beta\omega_{ho}^2/\pi$
where $\omega_{ho}=(\widetilde{\omega}_x\widetilde{\omega}_y)^{1/2}$. The condition of self-consistency imposes that the parameter $\eta_0$ is a solution of a third order equation:
\be
\eta_0^3+(1-2\Omega_i^2)\eta_0+\varepsilon\Omega_i=0.
\ee 
The time-dependent solution resulting from the sudden change of $\Omega$ is obtained by inserting the ansatz (\ref{ansatz}) into the hydrodynamic equations (\ref{continuity}) and (\ref{Euler}) 
\begin{eqnarray}
\dot \eta & = &  -(\alpha_x+\alpha_y)\eta-\Omega_{\rm fin}(\alpha_x-\alpha_y)-\beta a_{xy},\nonumber \\
\dot \alpha_x & = &  -\alpha_x^2-\eta^2-1-\varepsilon-2\beta a_x+2\Omega_{\rm fin}\eta,  \nonumber \\
\dot \alpha_y & = &  -\alpha_y^2-\eta^2-1+\varepsilon-2\beta a_y-2\Omega_{\rm fin}\eta,  \nonumber \\
\dot a_0 & = & - (\alpha_x+\alpha_y)a_0, \nonumber \\
\dot a_x & = &  -(3\alpha_x+\alpha_y)a_x-(\eta-\Omega_{\rm fin})a_{xy},\nonumber \\
\dot a_y & = & -(3\alpha_y+\alpha_x)a_y-(\eta+\Omega_{\rm fin})a_{xy}, \nonumber \\
\dot a_{xy} & = &  -2(\alpha_x+\alpha_y)a_{xy}-2(\eta+\Omega_{\rm fin})a_x  \nonumber \\
 & - & 2(\eta-\Omega_{\rm fin})a_y. 
\end{eqnarray}


\begin{thebibliography}{99}

\bibitem{heliumII}
R. J. Donnelly, \textit{Quantized Vortices in Helium II} (Cambridge University Press, Cambridge, England, 1991).

\bibitem{CornellImprinting}
M. R. Matthews, B. P. Anderson, P. C. Haljan, D. S. Hall, C. E. Wieman, E. A. Cornell, Phys. Rev. Lett. \textbf{83}, 2498 (1999). 

\bibitem{Dalibard1}
K. W. Madison, F. Chevy, W. Wohlleben, and J. Dalibard, Phys. Rev. Lett. \textbf{84}, 806 (2000). 

\bibitem{ScienceKetterle}
J. R. Abo-Shaeer, C. Raman, J. M. Vogels, W. Ketterle, Science \textbf{292}, 476 (2001).

\bibitem{Dalibard2}
K. W. Madison, F. Chevy, V. Bretin, and J. Dalibard, Phys. Rev. Lett. \textbf{86}, 4443 (2001). 

\bibitem{Cornell1}
P. C. Haljan, I. Coddington, P. Engels, and E. A. Cornell, Phys. Rev. Lett. \textbf{87}, 210403 (2001).

\bibitem{Foot}
E. Hodby, G. Hechenblaikner, S. A. Hopkins, O. M. Marag\`o, and C. J. Foot, Phys. Rev. Lett. \textbf{88}, 010405 (2001).

\bibitem{Dalibard3}
P. Rosenbusch, D. S. Petrov, S. Sinha, F. Chevy, V. Bretin, Y. Castin, G. Shlyapnikov, and J. Dalibard, Phys. Rev. Lett. \textbf{88}, 250403 (2002). 

\bibitem{KetterleMAgneticField}
Y. Shin, M. Saba, M. Vengalattore, T. A. Pasquini, C. Sanner, A. E. Leanhardt, M. Prentiss, D. E. Pritchard, and W. Ketterle, Phys. Rev. Lett. \textbf{93}, 160406 (2004). 

\bibitem{Spielman}
Y.-J. Lin, R. L. Compton, K. Jim\'enez-García, J. V. Porto and I. B. Spielman, Nature \textbf{462}, 628 (2009).

\bibitem{a1} D. L. Feder and C. W. Clark, Phys. Rev. Lett. \textbf{87}, 190401 (2001).

\bibitem{a2} A.L. Fetter, Phys. Rev. A \textbf{64}, 063608 (2001).

\bibitem{a3} K. Kasamatsu, M. Tsubota, and M. Ueda, Phys. Rev. A \textbf{66}, 053606 (2002).

\bibitem{a4} E. Lundh, Phys. Rev. A \textbf{65}, 043604 (2002).

\bibitem{a5} G. M. Kavoulakis and G. Baym, New J. Phys. \textbf{5}, 51 (2003). 

\bibitem{a6} U. R. Fischer and G. Baym, Phys. Rev. Lett. \textbf{90}, 140402 (2003).

\bibitem{a7} A. L. Fetter, Phys. Rev. A \textbf{68}, 063617 (2003).

\bibitem{a8} A. Aftalion and I. Danaila, Phys. Rev. A \textbf{69}, 033608 (2004). 

\bibitem{a9} A. D. Jackson, G. M. Kavoulakis, and E. Lundh, Phys. Rev. A \textbf{69}, 053619 (2004).

\bibitem{b1} A. D. Jackson and G. M. Kavoulakis, Phys. Rev. A \textbf{70}, 023601 (2004).

\bibitem{b2} A. Aftalion and I. Danaila, Phys. Rev. A  \textbf{69}, 033608 (2004). 

\bibitem{b3} A. L.  Fetter, N. Jackson, and S. Stringari, Phys. Rev. A \textbf{71}, 013605 (2005).

\bibitem{b4} A. Aftalion, X. Blanc, and J. Dalibard, Phys. Rev. A  \textbf{71}, 023611 (2005).

\bibitem{b5} A.L. Fetter, B. Jackson, and S. Stringari, Phys. Rev. A \textbf{71}, 013605 (2005).

\bibitem{b6} A. Aftalion, X. Blanc, and F. Nier, Phys. Rev. A  \textbf{73}, 011601(R) (2006). 

\bibitem{b7} H. Fu and E. Zaremba,  Phys. Rev. A \textbf{73}, 013614 (2006).

\bibitem{b8} G. Watanabe, S. A. Gifford, G. Baym, and C. J. Pethick Phys. Rev. A  \textbf{74}, 063621 (2006).

\bibitem{b9} A. Aftalion, X. Blanc, and N. Lerner, Phys. Rev. A  \textbf{79}, 011603(R) (2009).

\bibitem{b10} A. L. Fetter, Rev. Mod. Phys. \textbf{81}, 647 (2009).

\bibitem{brown_geonium_1986} L. S. Brown and G. Gabrielse, Rev. Mod. Phys. \textbf{58}, 233 (1986). 

\bibitem{castin} S. Sinha and Y. Castin, Phys. Rev. Lett. \textbf{87}, 190402 (2001). 

\bibitem{lobo}
C. Lobo, A. Sinatra, and Y. Castin, Phys. Rev. Lett. \textbf{92}, 020403 (2004).


\bibitem{GPELab}
X. Antoine and R. Duboscq, Computer Physics Communications \textbf{185}, 2969 (2014); \emph{ibid}  \textbf{193}, 95 (2015).

\bibitem{besse2004relaxation}
C. Besse, SIAM J. Numer. Anal, \textbf{42}, 934 (2004).

\bibitem{bao_ground_2005} W. Bao, P. A. Markowich, and H. Wang, Commun. Math. Sci. \textbf{3}, 57 (2005). 

\bibitem{guery-odelin_collective_1999}
D. Gu\'ery-Odelin, F. Zambelli, J. Dalibard and S. Stringari,  Phys. Rev. A \textbf{60}, 4851 (1999).

\bibitem{guery-odelin_spinning_2000}
D. Gu\'ery-Odelin, Phys. Rev. A \textbf{62}, 033607 (2000).

\bibitem{recati_overcritical_2001} A. Recati, F. Zambelli, and S. Stringari, Phys. Rev. Lett. \textbf{86}, 377 (2001).

\bibitem{dgocct} C. Cohen-Tannoudji, and D. Gu\'ery-Odelin, {\emph Advances in Atomic Physics: An Overview}, (Singapore, World Scientific, 2011).

\bibitem{Proukakis1}
N. G. Parker, N. P. Proukakis, C. F. Barenghi, and C. S. Adams, Phys. Rev. Lett. \textbf{92}, 160403 (2004).

\bibitem{Proukakis2}
N. G. Parker, N. P. Proukakis, and C. S. Adams, Phys. Rev. A \textbf{81}, 033606 (2010).


\bibitem{contrast}
The contrast is defined here as follows: we consider only the first oscillation of the curve over one period and compute the mean value $\overline{\Delta E/E_0}$. 
The contrast of the curve is then the difference between the mean and minimal values, divided by the difference between the maximal and mean values, over the considered period.

\bibitem{footnote2}
That is, we find a fidelity larger than $0.9999$. This is the reason why we use the term of quasi-reversibility.

\end{thebibliography}

\end{document}